\documentstyle[12pt]{article}
\def\bcn{\begin{center}}
\def\ecn{\end{center}}
\newcommand{\bea}{\begin{eqnarray}}
\newcommand{\eea}{\end{eqnarray}}
\newcommand{\beq}{\begin{equation}}
\newcommand{\eeq}{\end{equation}}

\newcommand{\sm}{standard model}

\newcommand{\xs}{cross section}
\newcommand{\br}{branching ratio}

\newcommand{\EW}{electroweak}

\newcommand{\cm}{center of mass}

\newcommand{\ep}{\mbox{$e^+e^-$}}

\def\lr3{$SU(3)_L\otimes SU(3)_R$}

\def\z0{$Z^0$}
\def\Z0{$Z^0$}

\def\ep{$e^+e^-$}

\def\cm{centre of mass}

\def\gsim{\buildrel{\lower.7ex\hbox{$>$}}\over{\lower.7ex\hbox{$\sim$}}}
\def\lsim{\buildrel{\lower.7ex\hbox{$<$}}\over{\lower.7ex\hbox{$\sim$}}}

\begin{document}
\thispagestyle{empty}
\setcounter{page}{0}

\begin{flushright}
MPI-PhT/94-54\\
PRL-TH-94/24\\
August 1994
\end{flushright}
\bigskip
\bigskip
\bigskip
\bigskip
\bigskip

\begin{center}
{\Large\bf Top Quarks and $CP$ Violation in Polarized $e^+e^-$ Collisions}\\

\bigskip\bigskip\bigskip\bigskip
\large
Frank Cuypers

{\it Max-Planck-Institut f\"ur Physik,
Werner-Heisenberg-Institut,}

{\it D--80805 M\"unchen, Germany}

{\tt cuypers@iws166.mppmu.mpg.de}

\bigskip\bigskip\bigskip
Saurabh D.~Rindani

{\it Theory Group, Physical Research Laboratory}

{\it Navrangpura, Ahmedabad 380009, India}

{\tt saurabh@prl.ernet.in}

\vspace{1.5ex}
\end{center}
\bigskip
\bigskip
\bigskip

\begin{abstract}
\noindent
Electroweak dipole moments of the top quark are conjectured
and their effect in polarized \ep\ collisions is examined
for the expectation values of two different $CP$-odd observables.
One of these observables probes the real part of the dipole moments
whereas the other probes their imaginary part.
It turns out
that varying the polarization of the electron beam
substantially enhances the resolving power of the experiment.
\end{abstract}
\newpage

Now that substantial evidence for the existence of the top quark
has finally been gathered \cite{CDF},
the exploration of its properties becomes an important endeavour.
It is not unnatural to expect these properties
to differ from those of the other known quarks,
since its mass is already substantially heavier.
The questions which
naturally arise are: How well are its couplings described by the
standard model? Is the top quark really elementary or is it a
composite of other more elementary objects?
The answers to these questions may be found
in studying the interactions of the top quark with
other particles and comparing them with predictions from the
standard model and other extended models, or analyzing them in a
model-independent manner. In particular, it would be interesting
to investigate whether top couplings can conserve $CP$, a symmetry so
far known to be violated only in the $K$-meson system.

Possible $CP$ violating couplings of fermions are electric
dipole type interactions with the electromagnetic field \cite{Leo},
and the analogous ``weak" dipole coupling to the $Z^0$ field.
These can arise at the one-loop level, for example, in certain
models of $CP$ violation like the
two-Higgs-doublet model or the minimal supersymmetric standard
model. They
would show up in the production of $t\overline{t}$ in $e^+e^-$
or hadronic colliders. It is our purpose to study the possibility
of measuring these
dipole type of couplings in $e^+e^-$ colliders, especially with
longitudinal beam polarization.
We do not restrict ourselves to any particular model, but
parametrize the $CP$ violation in terms of effective electric
and weak dipole form factors.

At a linear collider of the next generation
(CLIC, JLC, NLC, TESLA, VLEPP,\dots)
top quark pairs should be produced in sufficient abundance
(a few thousand events)
to allow simple studies of its nature.
The main production mechanism
proceeds at the Born level
by the $s$-channel annihilation of initial electron-positron pairs
into virtual photons or neutral weak gauge bosons,
and their subsequent splitting into top-antitop pairs
\beq
e^+ e^- \to \gamma^*,Z^{0*} \to t \bar t
{}~.\label{e1}
\eeq

There have been several suggestions to measure possible \EW\ dipole
moments of the top quark in top pair-production at $e^+e^-$ and hadronic
colliders. Various
experiments have been suggested to perform these measurements by
making use of $CP$-odd quantities. These include measuring the polarization
asymmetry \cite{Dono,Schmidt,Kane} of the $t\overline{t}$ pair through
the energy asymmetry of the charged leptons arising in their decays
\cite{Schmidt,Changtt}, as well as
the up-down asymmetry of these leptons with respect to the
production plane of $t\overline{t}$ \cite{Changtt},\footnote{The
effect of longitudinal beam polarization of these asymmetries is
studied in Ref.~\cite{Poulose}} and various
$CP$-odd momentum
correlations among the decay products of $t\overline{t}$ \cite{Bern1,Bern2}.
It has also been realized in the context of the measurement of
the $\tau$ dipole moment that longitudinal polarization of the
electron beam (and possibly also the positron beam) in the
$e^+e^-$ experiments can enhance certain momentum correlations
and can lead to enhanced sensitivity \cite{rin1}.
We investigate here the effect on similar correlations
of longitudinal beam polarization in the process
(\ref{e1}).

Quite apart from the actual numerical results we obtain, we will
see that the  main advantage of using polarized beams is that whereas in
the absence of polarization a
$CP$-odd correlation gives information on a certain combination
of the electric and weak dipole moments, its polarization
dependence can be used to get independent information on a
different combination. This helps to measure or obtain limits on
both types of dipole  moments. In the absence of polarization,
obtaining the same information would require measuring
at least two independent correlations.

It should be straightforward to obtain high degrees of
longitudinal polarization of the initial state electron beams
at a linear collider.
The same is not true, though,
for the positron beams.
This poses the problem that in experiments with only the
electron beam polarized, the initial state would not be a $CP$
eigenstate, and therefore $CP$-odd correlations are not
necessarily a measure of the $CP$ violation of the interaction.
However, if we neglect radiative corrections, only the
left-right and right-left combinations of electron and positron
helicities couple to the $\gamma$ and $Z^0$ in the
limit of vanishing electron mass. Hence the corrections, if any,
are tiny. Hard collinear emission of a photon from the initial
$e^-$ or $e^+$ can flip helicity, however, even in the limit of
vanishing electron mass \cite{FS}. This leads to non-zero $CP$-odd
correlations even in the absence of $CP$-invariant interactions,
and this background must be calculated and subtracted. While we
postpone a detailed calculation, we will argue that our
results are not seriously affected by this background.

\begin{figure}
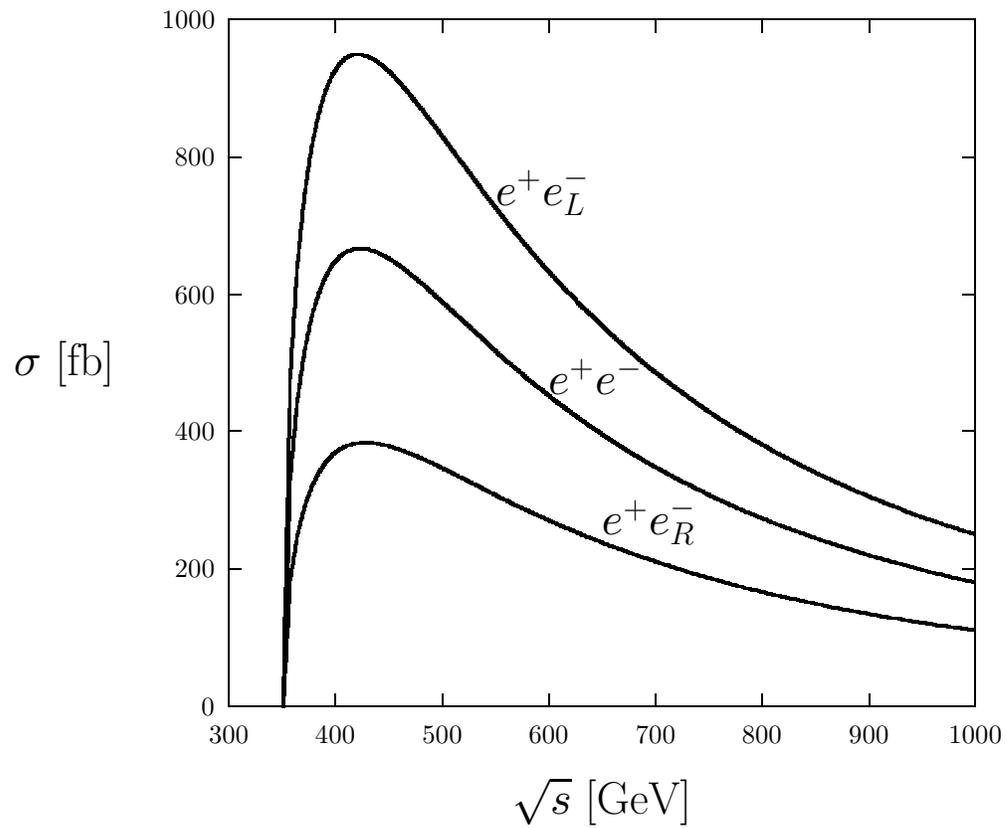

\input sig.tex
\caption{Energy dependence of the top quark pair-production cross section
in the presence and absence of polarization.
The top quark mass is taken to be 175 GeV.}
\label{f1}\end{figure}

At lowest order the polarized \xs\ for the process (\ref{e1}) is
\beq
\sigma(e^+e^- \to t\bar t) =
4\pi\alpha^2s ~ \sqrt{1-x} ~~ \Sigma
{}~,\label{e4}
\eeq
where
\beq
x = {4m_t^2\over s}
\label{e05}
\eeq
and
\bea
\Sigma & = &
	{1 \over s^2} ~ \left(1+{x\over 2}\right) ~
	{v_e^\gamma}^2{v_t^\gamma}^2 \nonumber\\
&+&	{2 \over s(s-m_Z^2)} ~ \left(1+{x\over 2}\right) ~
	v_e^\gamma v_t^\gamma ~ \left(v_e^Z-pa_e^Z\right) v_t^Z \nonumber\\
&+&	{1 \over (s-m_Z^2)^2} ~
	\left({v_e^Z}^2+{a_e^Z}^2-2pv_e^Za_e^Z\right) ~
	\left[ (1+{x\over2}) {v_t^Z}^2 + (1-x) {a_t^Z}^2 \right]
{}~.\label{e5}
\eea
As usual,
$s$ is the \cm\ energy squared,
$\theta_w$ is the weak mixing angle
taken in what follows to be given by $\sin^2\theta_w=.22$,
$\alpha=1/128$ is the fine structure constant and
$m_Z=91$ GeV and $m_t=175$ GeV are the $Z^0$ and top quark masses.
The latter value
has been chosen according to recent experimental results \cite{CDF,LEP}.
The degree of longitudinal polarization of the initial electron beam is $p$,
where $p=\pm1$ for right and left helicities respectively.
The electron and top vector and axial couplings
to the photon and the $Z^0$
are given by
\bea
v_e^\gamma = 1 &,& v_e^Z =
(1-4\sin^2\theta_w)/(4\sin\theta_w\cos\theta_w) \nonumber\\
a_e^\gamma = 0 &,& a_e^Z = 1/(4\sin\theta_w\cos\theta_w) \nonumber\\
v_t^\gamma = -{2\over3} &,& v_t^Z = (-1+{8\over3}\sin^2\theta_w)
/(4\sin\theta_w\cos\theta_w) \nonumber\\
a_t^\gamma = 0 &,& a_t^Z = -1/(4\sin\theta_w\cos\theta_w) \label{e9}~.
\eea
It is a straightforward task to incorporate the width of the $Z^0$.
The numerical effect, though,
is minute but has been taken into account in the numerical analysis.
The dependence of the top pair-production \xs\
on the \cm\ energy
is depicted in Fig.
\ref{f1}
for a fully left- and right-polarized electron beam
and in the absence of polarization.

The two top quarks produced in the reaction (\ref{e1})
subsequently decay into charged weak gauge bosons
and bottom quarks
with a \br\ which can for all practical purposes be taken equal to one:
\beq
t \to W^+ b
\qquad\qquad\qquad
\bar t \to W^- \bar b
{}~.\label{e2}
\eeq

Some information about the polarization of the final state top quarks
can also be inferred from the angular distributions of its decay products.
Indeed,
in the \cm\ frame of the (anti-) top quark,
the (anti-) bottom quark emerges with an angle $\theta$
with respect to the spin of its parent
which is distributed according to
\beq
{1\over\Gamma}{d\Gamma\over d\cos\theta}=
{1\over2} (1 \pm \beta \cos\theta)
{}~,\label{e6}
\eeq
where the mass of the bottom quark is neglected and
\beq
\beta = {m_t^2 - 2m_W^2 \over m_t^2 + 2m_W^2}
{}~.\label{e7}
\eeq

In the presence of an \EW\ dipole moment of the top quark,
the \sm\ lagrangian has to be supplemented with the following terms:
\beq
{\cal L}_{EWDM} =
-{i\over2} ~ d^V_t ~
[\bar t \sigma^{\mu\nu} \gamma_5 t] ~
(\partial_\mu V_\nu - \partial_\nu V_\mu)
\qquad\qquad
(V=\gamma,Z)
{}~,\label{e3}
\eeq
where $d^\gamma_t$ and $d^Z_t$
are the magnetic and weak dipole moments of the top quark.
It should be noted at this stage that these moments are by no means constants,
but rather energy dependent form factors.

In the following we concentrate on the two $CP$-odd observables \cite{rin1}
\bea
O_1 &=& (\vec p_b \times \vec p_{\bar b}) \cdot \vec 1_z  \label{e10}\\
O_2 &=& (\vec p_b + \vec p_{\bar b}) \cdot \vec 1_z  \label{e11}~,
\eea
where $\vec 1_z$ is the unit vector aligned with the incoming positron beam.
In the presence of \EW\ dipole moments (\ref{e3}),
these observables aquire non-vanishing expectation values. These
are not sensitive to possible $CP$ violation in top decay, and
we have therefore ignored it.
Neglecting the higher order terms in the dipole moments $d_t^{\gamma,Z}$,
we find
\clearpage
\bea
\langle O_1\rangle
&=&
{sm_t\over12}~(1-x)~\epsilon^2~\beta~\Sigma^{-1} \nonumber\\
&\Big\{&
{1\over s^2} ~ C^{\gamma\gamma} ~ {v_e^\gamma}^2 ~
v_t^\gamma ~ {\cal R}e~d_t^\gamma \nonumber\\
&+&
{1\over s(s-m_Z^2)} ~ C^{Z\gamma} ~ v_e^\gamma v_e^Z ~
(v_t^Z-{\beta\over3}a_t^Z) ~ {\cal R}e~d_t^\gamma \nonumber\\
&+&
{1\over s(s-m_Z^2)} ~ C^{Z\gamma} ~ v_e^\gamma v_e^Z ~
v_t^\gamma ~ {\cal R}e~d_t^Z \nonumber\\
&+&
{1\over (s-m_Z^2)^2} ~ C^{ZZ} ~ ({v_e^Z}^2+{a_e^Z}^2) ~
(v_t^Z-{\beta\over3}a_t^Z) ~ {\cal R}e~d_t^Z
\quad \Big\}
\label{e15} \\
\langle O_2\rangle
&=&
{\sqrt{s}m_t\over3}~(1-x)~\epsilon~\beta~\Sigma^{-1} \nonumber\\
&\Big\{&
{1\over s(s-m_Z^2)} ~ C^{Z\gamma} ~ v_e^\gamma v_e^Z ~
a_t^Z ~ {\cal I}m~d_t^\gamma \nonumber\\
&+&
{1\over (s-m_Z^2)^2} ~ C^{ZZ} ~ ({v_e^Z}^2+{a_e^Z}^2) ~
a_t^Z ~ {\cal I}m~d_t^Z
\quad \Big\}
{}~,\label{e16}
\eea
where
\beq
\epsilon = 1 - {m_W^2 \over m_t^2}
\label{e31}
\eeq
and
\bea
C^{\gamma\gamma} &=& -p \nonumber\\
C^{\gamma Z} &=& {a_e^Z \over v_e^Z}-p \nonumber\\
C^{ZZ} &=& {2v_e^Za_e^Z \over {v_e^Z}^2+{a_e^Z}^2}-p ~.\label{e32}
\eea
Note that $O_1$ being $CPT$-even
it is sensitive to the real parts of the dipole moments.
In contrast,
$O_2$ is $CPT$-odd
and provides thus a measure of the imaginary parts of the moments.

If a non-vanishing average values $\langle O\rangle $ is observed,
it has a statistical significance only as far as
it is compared with the expected variance $\langle O^2\rangle $.
For instance,
to observe a deviation from the \sm\ expectation
with better than 3 standard deviations
({\em i.e.}\ 99.7\%\ confidence)
one needs
\bea
\langle O\rangle  &\ge& 3\sqrt{{\langle O^2\rangle  \over n_{t\bar t}}}
{}~,\label{e40}
\eea
where $n_{t\bar t} = {\cal L}\sigma(e^+e^- \to t\bar t)$
is the number of events and $\cal L$ is the collider luminosity.

\clearpage
We find for the variances
\bea
\langle O^2_1\rangle
&=&
{sm_t^2\over2880}~\epsilon^4~\Sigma^{-1} \nonumber\\
&\Big\{&
{1\over s^2} ~ {v_e^\gamma}^2 ~
\left[ {v_t^\gamma}^2 ~ \left(24+2x-11x^2+4\beta^2(1-x)^2\right) \right]
\nonumber\\
&+&
{2\over s(s-m_Z^2)} ~ v_e^\gamma v_e^Z ~
\Big[ v_t^\gamma v_t^Z ~ \left(24+2x-11x^2+4\beta^2(1-x)^2\right) \nonumber\\
&&\qquad	- v_t^\gamma a_t^Z ~ 2(1-x)(6-x)\beta \Big] \nonumber\\
&+&
{1\over (s-m_Z^2)^2} ~ ({v_e^Z}^2+{a_e^Z}^2) ~
\Big[ {v_t^Z}^2 ~ \left(24+2x-11x^2+4\beta^2(1-x)^2\right) \nonumber\\
&&\qquad	{a_t^Z}^2 ~ \left(24-14x-4\beta^2(1-x)\right)(1-x) \nonumber\\
&&\qquad	- v_t^Z a_t^Z ~ 4(1-x)(6-x)\beta \Big]
\quad \Big\}
\label{e41} \\
\langle O^2_2\rangle
&=&
{s\over720}~\epsilon^2~\Sigma^{-1} \nonumber\\
&\Big\{&
{1\over s^2} ~ {v_e^\gamma}^2 ~
\left[ {v_t^\gamma}^2 ~ (4+7x+4x^2)(3-\beta^2) \right] \nonumber\\
&+&
{2\over s(s-m_Z^2)} ~ v_e^\gamma v_e^Z ~
\left[ v_t^\gamma v_t^Z ~ (4+7x+4x^2)(3-\beta^2) \right] \nonumber\\
&+&
{1\over (s-m_Z^2)^2} ~ ({v_e^Z}^2+{a_e^Z}^2) ~
\Big[ {v_t^Z}^2 ~ (4+7x+4x^2)(3-\beta^2) \nonumber\\
&&\qquad
+	{a_t^Z}^2 ~ \left(12+18x-4\beta^2(1-x)\right)(1-x) \Big]
\quad \Big\}
{}~.\label{e42}
\eea

Implementing Eq.~(\ref{e40})
one obtains the areas in the $(d_t^\gamma,d_t^Z)$ plane
which cannot be explored with sufficient confidence.
Clearly,
because of the linear dependence
of the expectation values (\ref{e15},\ref{e16}) on the dipole moments,
these areas are delimited by straight lines which are equidistant from
the \sm\ expectation
$d_t^\gamma = d_t^Z = 0$.
The slope of these straight lines
varies with the polarization of the initial electron beam,
as is shown in Figs~\ref{f2} and \ref{f3}.
For fully polarized beams the bands are narrowest.
As the degree of polarization is decreased,
the bands rotate around fixed points and become wider.
The results of Figs~\ref{f2} and \ref{f3}
have been obtained assuming a \cm\ energy of 500 GeV,
a luminosity of 10 fb$^{-1}$
and an overall conservative $b$- and $W$-tagging efficiency of 10\%.
Note that if the tagging efficiencies were to be improved
without loss of purity
from 10\%\ to 50\%\ or 90\%,
the bounds on the dipole moments in Figs~\ref{f2}\ and \ref{f3}\
would shrink by a factor of .45 or .33.

It is clear from Figs~\ref{f2} and \ref{f3} that in the absence
of polarization, the correlations $\langle O_1 \rangle$ and
$\langle O_2 \rangle $ are extremely
insensitive to ${\cal R}e~d^{\gamma}_t$ and ${\cal
I}m~d^{Z^0}_t$ respectively. However, the inclusion of
polarization makes the correlations sensitive to real as well as
imaginary parts of both type of dipole moments, due to the
rotation of the bands described above.
Note also that a single measurement
(with or without polarization)
cannot exclude large dipole moments:
in some unfortunate situations,
the electric and weak dipoles
can assume large values,
but their effects cancel out
so that no $CP$ violation is apparent.
However,
if the information from two measurements with opposite electron polarization
is combined,
only small values
(in the vicinity of the origin)
of the \EW\ dipole moments can escape detection.

Although the top pair-production \xs\ drops rapidly beyond 420 GeV,
the effect of the \EW\ dipole moments keeps augmenting
with increasing energies.
This can easily be traced back to the higher dimension
of the operators (\ref{e3}).
As can easily be inferred from the asymptotic behaviours
of $\langle O\rangle $, $\langle O^2\rangle $ and $n_{t\bar t}$,
the sensitivity of $O_1$ saturates while the sensitivity of $O_2$ is bound
to decrease beyond a certain energy,
which turns out to be around 750 GeV.
At this energy
the sensitivities of $O_1$ and $O_2$ improve by 60\%\ and 30\%\ respectively.
At higher energies
the sensitivity of $O_1$ almost doubles.
It should, however, be kept in mind
that the dipole moments being actually energy dependent form factors,
their magnitude is expected to decrease rapidly
at energies exceeding the scale of ``new physics''.

We now examine to what extent collinear helicity-flip photon
emission from the initial state affects our analysis. First of
all, there can be no $CP$-conserving background at order
$\alpha$ from such a
process for $\langle O_1\rangle $, since $O_1$ is $T$-odd. Any $CP$-conserving
process can only contribute to it if the amplitude has an
absorptive part, and to order $\alpha$, initial-state photon
emission occurs only at tree-level.
In contrast,
$\langle O_2\rangle$ can get such
a contribution, since $O_2$ is $T$ even.
In principle this \sm\ contribution to the signal can be computed
and subtracted from the values of $\langle O_2\rangle$ (\ref{e16}).
However,
since the helicity-flip photon spectrum is hard \cite{FS},
a cut on the final-state energy requiring it to be larger than
$(1-\delta) \sqrt{s}$ would suppress this background correlation by a
factor of at least $\alpha /(2\pi)\delta^3$.
Even an easily implementable cut with $\delta=0.05$
would thus suffice to render this background harmless.
Of course,
this would mean restricting the analysis
to hadronic decay modes of the $W^\pm$.
However,
the loss in efficiency is not dramatic
and can be considered as being included in the aforementioned
10\%\ overall $b$- and $W$-tagging efficiency.

We have compared our results
with the sensitivities obtained
by studying energy and up-down asymmetries of leptons
arising in top decay \cite{Changtt}
or correlations in the absence of polarization \cite{Bern1}.
We find that under similar assumptions
the polarized experiment we suggest
can improve the observability limits on the \EW\ dipole moments
by more than an order of magnitude.

To conclude, we have studied the role of longitudinal
polarization of electron beams in high energy $e^+e^-$ collisions
for measuring the real and imaginary parts of the top quark electric
and weak dipole moments by determining correlations of certain $CP$-odd
observables. We have calculated the correlations as well as the
variances of the observables using analytical expressions, and
obtained the sensitivities of these observables to the dipole
moments at a linear collider operating at $\sqrt{s}=500$~GeV and with an
integrated luminosity of 10 fb$^{-1}$. We find that the
sensitivities for the \EW\ dipole moments are greatly enhanced in
the presence of polarized electron beams.
The real and imaginary parts of the dipole moments
can be probed at the 3-$\sigma$ level
down to values below 1 e atto-m.

\clearpage

\clearpage
\begin{figure}
\input re.tex
\caption{Bounds from $\langle O_1\rangle $ on ${\cal R}e~d_t^{\gamma,Z}$
[e am]
according to Eq.~(\protect\ref{e40}).}
\label{f2}
\end{figure}

\clearpage
\begin{figure}
\input im.tex
\caption{Bounds from $\langle O_2\rangle $ on ${\cal I}m~d_t^{\gamma,Z}$
[e am]
according to Eq.~(\protect\ref{e40}).}
\label{f3}
\end{figure}


\begin{thebibliography}{rin1}
\bibitem{CDF} CDF Collaboration, F.~Abe {\it et al.}, Phys.~Rev.~
Lett.~{\bf 73}, 225 (1994); Phys.~Rev.~D (to appear).
\bibitem{Leo} F.~Hoogeveen and L.~Stodolsky, Phys.~Lett.~B {\bf 212}, 505
(1988).
\bibitem{Dono} J.F.~Donoghue and G.~Valencia, Phys.~Rev.~
Lett.~{\bf 58}, 451 (1987).
\bibitem{Schmidt} C.R.~Schmidt and M.E.~Peskin, Phys.~Rev.~Lett.~
{\bf 69}, 410 (1992).
\bibitem{Kane} G.L.~Kane, G.A.~Ladinsky and C.-P.~Yuan, Phys.~Rev.~D
{\bf 45}, 124 (1991).
\bibitem{Changtt} D.~Chang, W.-Y.~Keung and I.~Phillips, Nucl.~Phys.~B {\bf
408}, 286 (1993).
\bibitem{Poulose} P.~Poulose and S.D.~Rindani, in preparation.
\bibitem{Bern1} W.~Bernreuther, T.~Schr\"oder and T.N.~Pham,
Phys.~Lett.~B {\bf 279}, 389 (1992); W.~Bernreuther and P.~
Overmann, Nucl.~Phys.~B {\bf 388}, 53 (1992), Heidelberg
preprint HD-THEP-93-11 (1993).
\bibitem{Bern2} W.~Bernreuther and
A.~Brandenburg, Phys.~Lett.~B {\bf 314}, 104 (1993); J.P.~
Ma and A.~Brandenburg, Z.~Phys.~C {\bf 56}, 97 (1992); A.~
Brandenburg and J.P.~Ma, Phys.~Lett.~B {\bf 298}, 211 (1993), W.~
Bernreuther and A.~Brandenburg, Phys.~Rev.~D {\bf 49}, 4481 (1994).
\bibitem{rin1}{B.~Ananthanaryan, S.D.~Rindani, PRL-TH-93/17 (to
appear in Phys.~Rev.~Lett.); PRL-TH-94/7 ( to appear in
Phys.~Rev.~D).}
\bibitem{FS} B.~Falk, L.M.~Sehgal, Phys.~Lett.~B {\bf 325} (1994) 509.
\bibitem{LEP}{The LEP Collaborations, Phys.~Lett.~B {\bf 276}, 247 (1992);
CERN preprint PPE/93-157.}
\end{thebibliography}
\end{document}